\documentstyle[12pt,epsfig]{article}

\begin{document}
\newcommand \be
{\begin{equation}}
\newcommand \ee {\end{equation}}
\newcommand \ba
{\begin{eqnarray}}
\newcommand \ea
{\end{eqnarray}}
\newcommand{\siml}{\stackrel{<}{\sim}}
\newcommand{\subs}[1
]{{\mbox{\scriptsize #1}}}

\title{Financial markets as adaptive ecosystems}

\author{Marc Potters$^\ast$, Rama Cont$^{\ast, \#}$  
        and
Jean-Philippe Bouchaud$^{\dag,\ast}$\\[0.5em]
  {\small $^\ast$Science \&
Finance, 109--111 rue Victor Hugo,}\\
  {\small  92532 Levallois Cedex,
France}\\[0.5em]
  {\small $^{\dag}$Service de Physique de l'\'Etat
Condens\'e,
               Centre d'\'etudes de Saclay,} \\ 
  {\small Orme des
Merisiers, 91191 Gif-sur-Yvette Cedex, France}
\\[0.5em]
{\small $^{\#}$Laboratoire de Physique de la Mati\`ere Condens\'ee CNRS URA 190}\\
{\small Universit\'e de Nice- Sophia Antipolis B.P. 70, 06108 Nice, France}\\
}

\maketitle

\begin{center}
First version: Sept. 1996. This version: June 1997.
\end{center}

\begin{center}
02.50  Probability theory, stochastic processes and statistics\\
05.40.+j  Fluctuation phenomena, stochastic processes and statistics\\
89.90+n Other areas of general interest to physicists
\end{center}

\begin{abstract} 

We show, by studying in detail the market prices of options on liquid markets, that
the market has empirically corrected the simple, but inadequate
Black-Scholes formula to account for two important statistical
features of asset fluctuations: `fat tails' and correlations in the scale of fluctuations. These aspects, although not included in the pricing models, are very precisely reflected in the price fixed by the market as a whole. Financial markets thus behave as rather efficient adaptive systems. 

\end{abstract}

Options
markets offer an interesting example of the adaptation of a
population
(the traders) to a complex environment, through trial and errors
and
natural selection (inefficient traders disappear quickly). The problem
is the
following: an `option' is an insurance contract protecting its owner
against the rise
(or fall) of financial assets, such as stocks, currencies,
etc. The problem of
knowing the value of such contracts has become
extremely acute ever since organized option markets
opened twenty five years ago,
allowing one to buy or sell options much like stocks. 
Almost
simultaneously, Black and Scholes (BS) proposed their famous option
pricing
theory, based on a simplified model for stock fluctuations, namely
the (geometrical) continuous time 
Brownian motion model. The most important parameter of
the model is the 
`volatility' $\sigma$, which is the standard deviation of
the market price's fluctuations. 
The Black-Scholes model is known to be based
on unrealistic assumptions but is nevertheless used 
as a benchmark by market participants. 
Guided by the Black-Scholes
theory, but constrained by the fact that
`bad' prices lead to arbitrage
opportunities, the option market fixes prices which
are close, but
significantly and systematically different from the BS
formula.
Surprisingly, a detailed study of the observed market prices clearly
shows
that, despite the lack of an appropriate model, traders have
empirically adapted to
incorporate some subtle information on the real
statistics of price changes. Although this ability to price financial 
assets correctly is often assumed in the literature (the `efficient market'
hypothesis), it is in general difficult to assess quantitatively, because
the `true' value of a stock, if it exists, is difficult to determine. The 
case of option markets is interesting in that respect, because the `true'
value of an option is, in principle, calculable.

More precisely, a `call' option is such that
if the price $x(T)$ of a given asset at
time $T$ (the `maturity') exceeds a
certain level $x_s$ (the `strike' price), the
owner of the option receives
the difference $x(T)-x_s$. Conversely, if $x(T) < x_s$,
the contract is
lost. To make a long story short \cite{BS,Hull,BS94,Book}, if $T$ is
small
enough (a few months) so that interest rate effects and average returns are negligible compared to
fluctuations, the `fair' price $\cal C$ of the option today ($T=0$),
knowing that the
price of the asset now is $x_0$ is simply given by \cite{Rq1}:
\be
{\cal
C}(x_0,x_s,T) = \int_{x_s}^\infty dx' \ (x'-x_s)
P(x',T|x_0,0)\label{price}
\ee
where $P(x',T|x_0,0)$ is the conditional
probability density that the stock price at time
$T$ will be equal to $x'$,
knowing its present value is $x_0$. Eq. (\ref{price})
means that the option
price is such that on average, there is no winning party.
Pricing correctly
an option is thus tantamount to having a good model for the
probability density $P(x',T|x_0,0)$.

There is fairly strong evidence that beyond a
time scale $\tau$ of the order of
ten minutes, the fluctuations of prices in liquid markets are
uncorrelated, but {\it not
independent} variables \cite{granger,Olsen,comment,houches,scaling}. In particular, it has been observed that
although the signs of successive price movements seem to
be independent, their magnitude - as represented by the absolute value
or square of the price increments- is correlated in time \cite{granger, scaling}: this is related
to the so-called `volatility clustering' effect \cite{ARCH,Olsen}.
More precisely one can represent the price $x(T)$ of the asset as 
\be x(T)=x_0 +
\sum_{k=0}^{\frac{T}{\tau}-1} \delta x_k
\ee
where the increments $\delta x_k$ are obtained as the {\it product of two
random variables}:

\be
\delta x_k = \epsilon_k \gamma_k
\ee
where $(\epsilon_k)_{k\geq 0}$ is a sequence of independent, identically distributed random variables of mean zero and unit variance, and $\gamma_k$ is a stochastic {\it scale factor} independent from the $\epsilon_k$s.
The sequence $(\gamma_k)_{k\geq 0}$ is considered to be a stationary random process but allowed to
 exhibit non-trivial correlations (see
below).  Under these hypotheses, the conditional distribution of $\delta x_k$, conditioned on
$\gamma_k$, may be written as:
\be
P(\delta x_k) \equiv
\frac{1}{\gamma_k} P_0\left(\frac{\delta x_k}{\gamma_k}\right)
\ee
where $P_0$ is independent of $k$. Models with conditionally Gaussian increments -- i.e. 
where $P_0$ is a Gaussian -- have been extensively studied 
\cite{ARCH} both in discrete time (ARCH models) and continuous time (stochastic
volatility models)
settings. The present model is more general since we do not assume that $P_0$ is Gaussian.

Let us first consider the case where $\gamma_k = \gamma_0$ is
independent of $k$, which
corresponds to the classical problem of a sum of independent, identically
distributed variables. Although $P(\delta x)$ is strongly non Gaussian (see, e.g. \cite{mantegna}), it has a finite variance \cite{houches} and the
Central Limit Theorem \cite{Feller} tells us that for large $N=T/\tau$, $P(x',T|x_0,0)$ will be close to a Gaussian. Using then Eq. (\ref{price}) essentially leads back to the BS formula \cite{Rq2}. For finite $N$, however,
there are corrections to the
Gaussian, and thus corrections to the BS price.
More precisely, the difference between $P(x',T|x_0,0)$ and the limiting
Gaussian distribution ${\cal G}_{x_0,\sigma^2}$ may be calculated using a {\em cumulant} expansion \cite{Feller}. To a very good approximation,
the
distribution $P_0(\delta x)$ is symmetric \cite{mantegna,Book}
for time scales less than a month i.e. drift effects are negligible
compared to fluctuations. This in turn implies that the 
third cumulant, which measures the skewness of the distribution, is
small, in which case the leading correction in the cumulant expansion
 mentioned above is, for large $N$, proportional to the {\it
kurtosis} $\kappa$, defined as $\kappa=\langle
\delta x^4 \rangle/\langle
\delta x^2 \rangle^2-3$ \cite{Feller}. $\kappa$ vanishes if the increments $\delta x$ are 
Gaussian random variables, and measures the `fatness' of
the tails of the distribution as compared
to a Gaussian. 

Neglecting higher order cumulants, the expansion takes the following form:

\be
\int_{-\infty}^{x} \left\{ P(x',T|x_0,0) - {\cal G}_{x_0,\sigma_T^2}(x')\right\} dx' = \frac{1}{\sqrt{2\pi}} e^{-u^2/2} \left[\frac{\kappa_T}{24} (u^3 - 3u ) + \ldots \right] 
\ee

where $u=(x-x_0)^2/\sigma_T^2$, $\sigma_T^2$ and $\kappa_T$ being the variance
and kurtosis corresponding to the scale $T$. ${\cal G}_{x_0,\sigma_T^2}$ is
the gaussian centered at $x_0$ of variance $\sigma_T^2$.

\begin{figure}
{\centering
\epsfig{figure=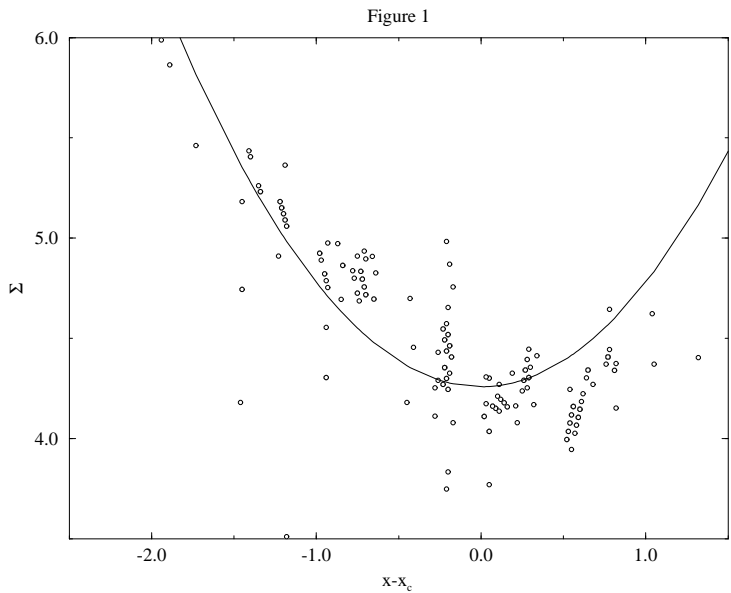,width=9cm,angle=270}}
\caption{Example of a smile curve: Implied volatility
$\Sigma(x_s,T)$ {\em vs}\/
distance from  strike price $(x-x_s)$ for a given $T$. The data
shown correspond to all 227
transactions of December options on the German Bund future 
(LIFFE) on November 13, 1995. This is a very
`liquid' market, meaning that
price anomalies are expected to be small, in particular for 
short maturities $T$.
Both call and put options are included
with put options transformed into call
options using the put-call
parity \protect\cite{Hull}. Volatilities are expressed as
annualized standard
deviation of price differences. According to Eq.
(\protect\ref{smile}) the data
should fall on a parabola. From a fit of the average curvature of
this parabola, we
extract the `implied kurtosis' $\kappa_{\subs{imp}}$ for a given
$N=
\protect\frac{T}{\tau}$. In this particular case we find $\kappa_\subs{imp}=1.92$
at
$N=144$ (9 trading days).
}
\label{fig1}
\end{figure}

It is then easy to
show, using Eq. (\ref{price}), that the leading correction to the BS
price can be reproduced by using
the BS formula, but with a modified value for
the volatility
$\sigma=\sqrt{\langle \delta x^2 \rangle}$ (which traders  call
the
`implied volatility' $\Sigma$), which depends both on the strike price
$x_s$
and on the maturity $T$ through: 
\be 
\Sigma(x_s,T) = \sigma
\left[1+ \frac{\kappa_T 
}{24}
\left(\frac{(x_s-x_0)^2}{\sigma_T^2}- 1
\right)\right] 
\label{smile}
\ee 
The fact that implied volatility depends on the strike price $x_s$ is known as the `smile effect',
because the plot of $\Sigma$ versus $x_s$, for a given value of $T=N\tau$ has
the shape of a smile (see Fig 1).

 That
the volatility had to be smiled up was realized long ago
by traders -- this
reflects the well known fact that the elementary increments have fat-tailed distributions: large fluctuations occur much more often than for a Gaussian random
walk. 

\begin{figure}
{\centering
\epsfig{figure=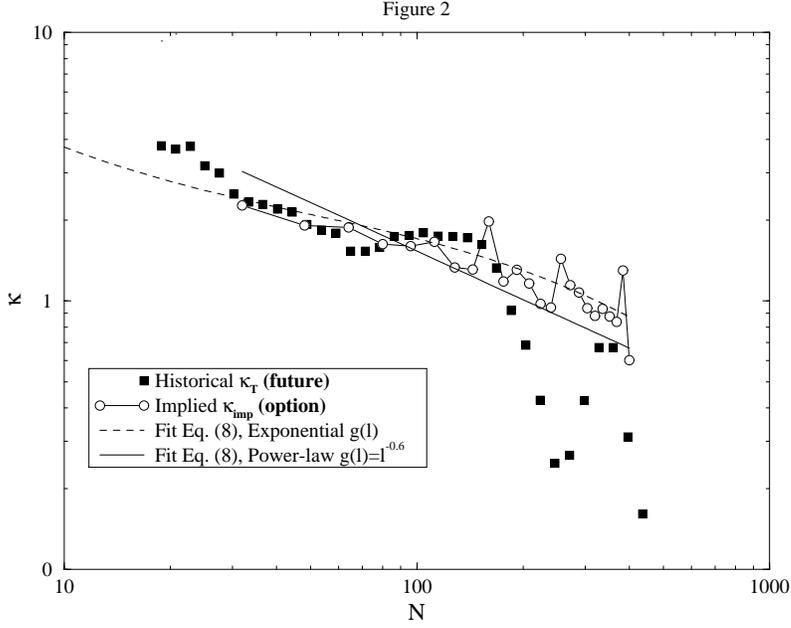,width=9cm,angle=270}}
\caption{Plot (in log-log coordinates) of the average implied kurtosis $\kappa_\subs{imp}$ (determined as in Fig. 1)
and of the empirical kurtosis $\kappa_T$ (determined directly
from the
historical movements of the Bund contract), as a function of the
reduced time
scale $N=T/\tau$, $\tau = 30$ minutes. All
transactions of options on the
Bund future from 1993 to 1995 were analyzed
along with 5 minute tick data of the
Bund future for the same period. The
growth of the error bars for the latter
quantity comes from the fact that
less data is available for larger $N$. Finally, we show for comparison a fit with
formula (\protect\ref{fit}), with $g(\ell) \simeq \ell^{-0.6}$, which leads to 
$\kappa_T \simeq T^{-0.6}$ (dark line). A fit with an exponentially decaying $g(\ell)$ is however also acceptable (dotted line).
}
\label{fig2}
\end{figure}

As shown in
Fig. 2, the smile formula (\ref{smile}) reproduces correctly the
observed
option prices on the `Bund' market provided the kurtosis
$\kappa_T$ in formula (\ref{smile})
becomes itself $T$-dependent. The shape of the `implied' kurtosis
$\kappa_\subs{imp}(T)$
as a function of $T$ is given in Fig. 2;
$\kappa_\subs{imp}(T)$ is seen to decrease more slowly than $T^{-1}$. 
However, if the increments $\delta x$
were independent and identically distributed (i.e. $\gamma_k
\equiv \gamma_0$), one
should observe that $\kappa_T = \kappa/N$.  

Let us then study directly the kurtosis of the distribution of the underlying
stock,
$P(x,T|x_0,0)$, as a function of $N\equiv T/\tau$. In Fig.
2, we have also shown
$\kappa_T$ as a function of $N$. One can notice that not only
$T\kappa_T$ is not constant (as it
should if $\delta x$ were identically
distributed), but actually
$\kappa_T$ matches quantitatively (at least for $N \leq 200$)  the
evolution of the
implied kurtosis $\kappa_\subs{imp}$! (Note that there is no adjustable
overall factor.) In other words, the price
over which
traders agree capture rather precisely the anomalous evolution
of
$\kappa_T$. A similar agreement has been found on other liquid 
option markets, where bid-ask spreads are sufficiently small to ascertain that
the quoted prices should indeed be set by a fair game condition. For `over the counter'
options, this is likely not to be the case, since a rather high risk premium is generally
included in the price.

As we shall show now, the non trivial behaviour
of $\kappa_T$ is related to the fact that the {\it scale} of the
fluctuations $\gamma_k$ is itself a time dependent random variable \cite{ARCH,granger}, with rather long range correlations. The random character
of $\gamma_k$ could come from 
the fact that $\gamma_k$ is related to the level of market activity, which fluctuates with time. 

We define the correlation function of the scale of fluctuations as:
\be
g(\ell)  =  \frac{\langle{\delta x_{k+\ell}^{2} \delta x_k^{2}}\rangle - \langle{\delta x_{k}^2}\rangle^2 }
{\langle{\delta x_{k}^{4}}\rangle - 
\langle{\delta x_{k}^2}\rangle^2 }
\ee
$g(\ell)$ is normalized such that $g(0)=1$. In this case, one can show
that Eq. (\ref{smile}) holds, with $\kappa_T$ given by:
\be
\kappa_T = \frac{\tau}{T} \left[
\kappa_\tau + 6 (\kappa_\tau +2) \sum_{\ell=1}^{N}
(1-\frac{\ell}{N})g(\ell)\right]\label{fit}
\ee
where $\kappa_\tau$ is the kurtosis of $\delta x = x(t+\tau) - x(t)$.
We have computed from historical data the correlation
function $g(\ell)$, which we show in Fig. 3. Interestingly,
$g(\ell)$ decreases rather slowly, as $\ell^{-\lambda}$, with 
$\lambda \simeq 0.6 \pm 0.1$, from minutes to several days. A similar decay of $g(\ell)$ was observed
on other markets as well, with rather close values for $\lambda$, such as the S\&P500 (for which $\lambda \simeq 0.37$) \cite{scaling} and the {\sc dem}/\$ market (for which $\lambda \simeq 0.57$). Remarkably,  Eq. (\ref{fit}) with $g(\ell) \propto \ell^{-0.6}$ leads to $\kappa_T \propto T^{-0.6}$, in good agreement with both the direct determination of $\kappa_T$
and the one deduced from the volatility smile, $\kappa_\subs{imp}$.
Note that the effect of a non zero kurtosis on Black-Scholes
prices was
previously investigated in \cite{kurt,corrado}. However, the 
relation between $\kappa_T$ and $\kappa_\subs{imp}$, and
their anomalous $T$ dependence, were not, to our knowledge, previously
reported.  

\begin{figure}
{\centering
\epsfig{figure=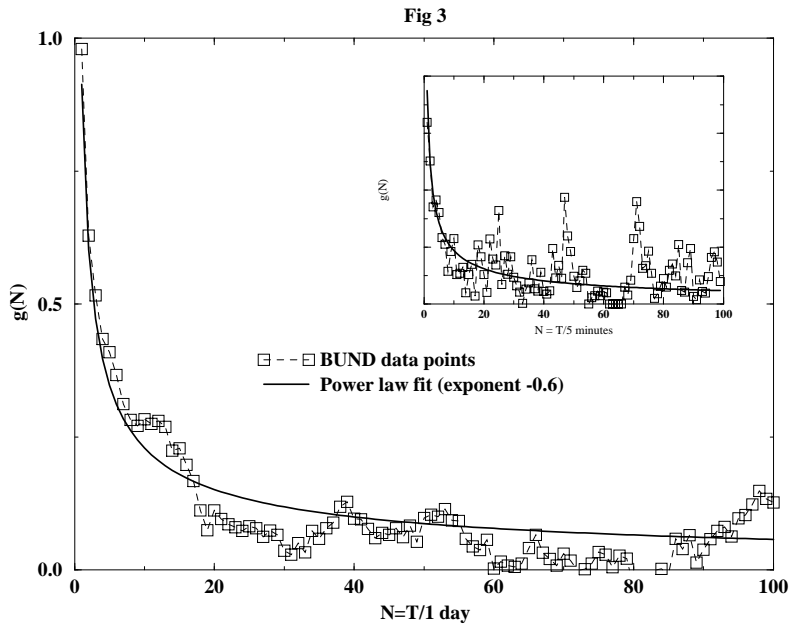,width=9cm,angle=270}}
\caption{Plot of the daily volatility correlation function $g(N)$ for the Bund future market, from 1991 to 1995. A fit by $g(N) \simeq N^{-\lambda}$ with $\lambda=0.6$ is shown for comparison. The same behaviour is found to persist for intra-day fluctuations (see Inset).
}
\label{fig3}
\end{figure}

In conclusion, we have shown by studying in detail the market
prices of options that
traders have evolved from the simple, but inadequate
BS formula to an empirical
know-how which encodes two important statistical
features of asset fluctuations: 
`fat tails' (i.e. a rather large kurtosis)
and  the fact that the scale of
fluctuations exhibits slowly decaying  correlations. 
These features, although not explicitly included in the theoretical pricing models used by traders, are nevertheless reflected rather precisely in the price fixed by the market as a whole. Option markets offer an interesting ground 
where `theoretical' and `experimental' prices can be systematically compared, and were found to agree rather well \cite{RqF}. This has enabled us to test quantitatively the idea that the trader population
behaves as an efficient adaptive system.


\begin{thebibliography}{99}
\bibitem{Rq2} Although
in principle BS use a log-normal, rather
than a normal distribution, the
difference is not relevant for the present discussion. See \cite{Book} for a
detailed discussion of this point.
\bibitem{RqF} Note that small systematic differences have in some cases been 
found. These differences are however within transaction costs.
\end{thebibliography}
\end{document}